# Quantitative portfolio selection: using density forecasting to find consistent portfolios

N. Meade[1], J.E. Beasley[2] and C.J. Adcock[3]

[1]Imperial College Business School, London SW7 2AZ, UK, n.meade@imperial.ac.uk
(Corresponding Author)
[2]Mathematics, Brunel University, Uxbridge UB8 3PH, UK, john.beasley@brunel.ac.uk
[3]School of Finance and Management, SOAS – University of London, London WC1H 0XG, UK, c.j.adcock@soas.ac.uk

**Abstract**

In the knowledge that the ex-post performance of Markowitz efficient portfolios is inferior to that implied ex-ante, we make two contributions to the portfolio selection literature. Firstly, we propose a methodology to identify the region of risk-expected return space where ex-post performance matches ex-ante estimates. Secondly, we extend ex-post efficient set mathematics to overcome the biases in the estimation of the ex-ante efficient frontier. A density forecasting approach is used to measure the accuracy of ex-ante estimates using the Berkowitz statistic, we develop this statistic to increase its sensitivity to changes in the data generating process. The area of risk-expected return space where the density forecasts are accurate, where ex-post performance matches ex-ante estimates, is termed the consistency region. Under the 'laboratory' conditions of a simulated multivariate normal data set, we compute the consistency region and the estimated ex-post frontier. Over different sample sizes used for estimation, the behaviour of the consistency region is shown to be both intuitively reasonable and to enclose the estimated ex-post frontier. Using actual data from the constituents of the US Dow Jones 30 index, we show that the size of the consistency region is time dependent and, in volatile conditions, may disappear. Using our development of the Berkowitz statistic, we demonstrate the superior performance of an investment strategy based on consistent rather than efficient portfolios.

## 1. Introduction

A well-known problem of quantitative portfolio selection is that mean-variance efficient portfolios fail to deliver the promise of their ex-ante expected return and variance. In this paper, we venture into the largely ignored and unexplored risk[1] -expected return space within the efficient frontier. Our objective is to identify portfolios whose ex-post performance corresponds to their ex-ante promise.

The efficient frontier is a set of portfolios expected to offer minimum risk for given levels of expected return, chosen from a specified universe of assets. Each efficient portfolio is found by a constrained minimisation of the variance of portfolio returns using historically based estimates of the means, variances and covariances of asset returns. It has long been known that the out-of-sample performance of these efficient portfolios frequently does not meet expectations. Michaud (1989) suggested that unless mean-variance optimisation was used with '*sophisticated adjustment of the inputs*', its use '*may often do more harm than good*'. Broadie (1993) expresses the trade-off in estimation: using too few historical observations leads to estimation errors; using too many historical observations means that the parameter estimates may be outdated. He concludes that '*points on the estimated frontier are optimistically biased predictors of actual portfolio performance*'. In our literature review we divide approaches to mean-variance portfolio selection into two paths: in Path One parameter estimation is made more robust; in Path Two constraints and objective functions are chosen to mitigate the impact of estimation errors.

This paper makes two contributions to the literature on portfolio selection.

Firstly, we propose an innovative empirical methodology, based on the measurement of density forecasting accuracy, to identify the region of risk-expected return space where portfolios will exhibit the same performance out-of-sample as they exhibited in-sample. For a universe of $N$ assets, we consider portfolios whose ex-ante performance lies within the risk-expected return space bounded above by the efficient frontier and bounded below by the return on the minimum variance portfolio. The efficient frontier is estimated using data from $M$ observations of asset returns. Over a series of $K$ time origins, the risk-expected return space containing the ex-post performance of the portfolios is computed. We identify the intersection of the ex-ante and ex-post risk-expected return spaces, in the sense that ex-post performance is consistent with ex-ante estimates. We call this intersection the consistency region and its upper boundary the consistency frontier. The methods of density forecasting are used to measure consistency. Specifically, we use the Berkowitz statistic which we extend in this paper to increase its sensitivity to a changing data generating process. The derivation of the consistency frontier places constraints on the historical performance of selected portfolios and thus is classified as a Path Two approach.

---

[1] In this paper the words risk and volatility are synonymous and are used interchangeably.

Secondly, we make a substantial extension to the ex-post efficient set mathematics described in Adcock (2013), who presented a derivation of the ex-post efficient frontier designed to overcome the biases in the ex-ante efficient frontier that result from the use of estimates. As this approach is concerned with the effect of parameter estimation ex-post, it clearly falls into Path One. The extension described below incorporates the effects of constraints and so spans both Paths.

We use density forecasting methods to examine the consistency of these ex-post efficient frontiers. We further examine the relative positions of the consistency frontier and the estimated ex-post frontier. These analyses allow a comparison between the empirical basis of the consistency frontier and the theoretical foundation of the estimated ex-post frontier.

In the first part of this analysis, we demonstrate that the consistency regions behave intuitively reasonably under the laboratory conditions represented by simulated multivariate normal data with a constant mean vector and covariance matrix; the only source of uncertainty being sampling error. Under the same conditions, we derive the ex-post frontier and discuss its position relative to the consistency region.

In the second part of this analysis, we use data from the constituents of the US Dow Jones 30 index to show how the consistency regions react to the changing dynamics of the market. In these non-laboratory conditions, uncertainty due to sampling error is exacerbated by a time dependent mean vector and covariance matrix. To improve the sensitivity of the consistency region to changing market conditions, we use a development of the Berkowitz statistic to better capture time dependency. To evaluate the economic implications of consistency, we compare investment strategies applied to efficient portfolios and to consistent portfolios.

This paper has eight sections; in the next section, we review the literature concerned with estimation errors in the efficient frontier. The background to accuracy measurement in density forecasting is given in Section 3, where we describe the Berkowitz statistic and propose an extension designed to increase its sensitivity to change. In Section 4, we describe the methodology for the construction of the consistent frontier. In Section 5, we summarise ex-post efficient set mathematics and describe extensions to this work. Using simulated data, we validate the models in Section 6. In Section 7, we describe the data set of Dow Jones 30 stocks and use these data to investigate the behaviour of the consistency region over time and explore the investment implications of portfolio selection based on the consistency frontier. Section 8 concludes.

## 2. Literature review

The wealth of research sparked by Markowitz's (1952) seminal work on portfolio selection is exemplified by the 2014 special issue of EJOR to celebrate the 60th Anniversary of his original paper. This research is reviewed by Kolm et al (2014), earlier surveys of work on mean-variance optimisation in portfolio selection include Fabozzi et al (2007) and Markowitz (2014). The usefulness of the efficient frontier has been questioned by many authors, for example Michaud (1989) and Broadie (1993). However, as Kolm et al (2014) comment these criticisms do not necessarily imply a flaw in the theory of risk-expected return optimisation, but do imply that a modified approach is needed in practice *"to achieve reliability, stability, and robustness with respect to model and estimation errors"*.

The portfolio selection problem considers a universe of *N* assets whose performance is summarised by a return vector $\boldsymbol{\mu}$ and a covariance matrix $\boldsymbol{\Sigma}$. Markowitz (1952) showed that quadratic programming can be used to establish an efficient frontier of portfolios of assets by finding the optimal set of weights on the assets, $\boldsymbol{\omega}$, such that the portfolios on the frontier have minimum risk for a given level of expected return. In practice, $\boldsymbol{\mu}$ and $\boldsymbol{\Sigma}$ are unknown and must be estimated from historical data. However, when the true (but unknown) parameters are replaced by sample estimates, $\hat{\boldsymbol{\mu}}_t$ and $\hat{\boldsymbol{\Sigma}}_t$, the search for a minimum variance portfolio is affected by errors in in-sample variances and covariances.

In his critique of mean-variance optimisation, Broadie (1993) suggests that improving the estimation of returns is the most effective way to reduce bias. Exploration of the problem of estimation bias has followed two paths. Path One looks at ways of making parameter estimation more robust. Path Two looks at modifying the constraints and the objective function to lessen the impact of estimation errors. These paths are not mutually exclusive. For example, improved estimates may be used with constraints or a modified objective function. Furthermore, the outcomes from the two paths may be similar, but this division is helpful for exposition.

One issue motivating Path One is that it has long been known that asset returns are not well described by the normal distribution; see for example, Blattberg and Gonedes (1974), Kon (1984), Barndorff-Nielson (1977) and a fuller survey in Meade (2010). Departures from normality and time dependency in mean and covariance add to the difficulties of selecting a portfolio whose ex-post performance reflects its ex-ante parameters. To mitigate the difficulties of estimation, robust estimation using Stein or shrinkage estimators (see James and Stein, 1961) was introduced by Jobson and Korkie (1980). Jorion (1986), Chopra, Hensel and Turner (1993), among others, explored this topic. In a survey of work on robust portfolios, Fabozzi, Huang and Zhou (2010) point out that the Bayesian approach is appropriate when the true data generating process is unknown. This paper is an update and companion to the comprehensive text by Fabozzi et al (2007). The Black-Litterman (1991) approach is an extension of the Bayesian approach to parameter estimation incorporating the prior views of investors, for more details see Fabozzi et al (2007). Jorion (1991) compares Bayes-Stein estimation with CAPM based estimation of mean asset returns. He finds that the latter method outperforms Bayes-Stein estimation which itself outperformed classical estimation. Michaud and Michaud (2007) show that the resampled efficient frontier proposed in Michaud (1998) is a Bayesian generalisation of the Markowitz solution. Ledoit and Wolf (2003) applied shrinkage estimation to the covariance matrix; they produce an optimally weighted average estimate using the sample covariance matrix and the single-index (CAPM) covariance matrix. Fabozzi, Huang and Zhou (2010) also point out that variance does not capture all components of what many investors consider as risk, but they further suggest that no single measure exists as a universal solution. In this context, Tu and Zhou (2004) devised an approach for the inclusion of uncertainty in the data generation process to account for fat tails in return density. Although this led to changes in portfolio weights, they concluded that the certainty-equivalent penalty for ignoring fat tails was low and that for a mean-variance investor the normality assumption worked well. Adcock (2014) describes a set of several multivariate skew-elliptical multivariate return probability distributions which lead to an extension of Markowitz's frontier, with an additional dimension for skewness, concluding that Markowitz's theory is robust to several departures from multivariate normality.

In Path Two, the concern is the mitigation of errors in estimation. Best and Grauer (1991a, 1991b) showed that small changes in expected returns lead to large changes in portfolio weights. Chopra & Ziemba (1993) considered estimation errors in expected returns as twice as important than errrors in the covariance matrix. Palczewski and Palczewski (2014) find that the relative importance of the estimation of covariance depends on the Sharpe ratio of the market portfolio and the frequency of observation of the data. DeMiguel, Garlappi and Uppal (2009) propose the most radical solution by ignoring all estimates and using equal weights for all assets, the so called naïve or 1/N approach.

Using a constraint-based-approach, Jagannathan and Ma (2003) motivate their analysis by drawing attention to the practical problem of selecting a portfolio from a universe of 500 assets. With a maximum of around 900 monthly observations of returns available there are fewer than four degrees of freedom for each parameter estimate. They focus on why the imposition of constraints helps reduce risk. If an asset has large estimated covariances with other assets, then it is likely to be drawn into the minimum variance portfolio (with the opposite signed weight to those on the other assets, assuming short-selling is allowed). Imposing a non-negativity constraint on the asset weight (i.e. not allowing short-selling) thus has the effect of reducing the covariances of assets bound by the constraint. Similarly, the imposition of upper bounds on weights tends to affect assets with low covariance with other assets, in this case effectively increasing some covariances. Indicating that their approach lies in the intersection between Path One and Path Two, Jagannathan and Ma (2003) note that the effect of these constraints is similar to using shrinkage in estimation.

Portfolio selection using robust optimisation is a well-studied approach that falls within Path Two. A set of possible parameter values called the uncertainty set is defined with the aim of finding the best performing portfolio under the worst possible values of the uncertainty set. In a thorough survey, Kim et al (2014) describe the range of objectives used to measure performance. Tütüncü and Koenig (2004) maximise the Sharpe ratio, other authors ignore returns and focus on minimising risk, for example Maillard et al (2010). Other objectives include: conditional value at risk, CVaR, see Rockafellar and Uryasev (2000) and Huang et al (2010); the Omega ratio (the expected returns above a given threshold divided by the expected returns lower than the threshold), see Kapsos et al (2014).

In the context of portfolio performance relative to that of competitors, Simões et al (2018) consider regret minimisation as an objective.

**Table 1. A summary of empirical comparisons between portfolio selection approaches**

| Authors | Conclusion | Data | Criterion |
|---|---|---|---|
| Frankfurter, Phillips and Seagle (1971) | Mean-variance portfolios are no more likely to be efficient than randomly selected portfolios. | Simulated multivariate normal data | Frequency of domination |
| Bloomfield, Leftwich and Long (1977) | Average return on tangency portfolio not significantly better than equal weighted portfolio. | US equities | Comparison of means |
| Jorion (1985) | Out-of-sample, shrinkage estimators significantly outperform sample mean. | 7 National indices, monthly (1972 – 1983) | Sharpe ratio |
| Broadie (1993) | 'points on the estimated frontier are optimistically biased predictors of actual portfolio performance' | Simulated data | Distances between 3 frontiers: using estimated parameters, true parameters and 'actual' frontier based on estimated weights with true parameters. |
| Chan, Karceski and Lakonishok (1999) | Out-of-sample performance of minimum variance portfolio dominates equal weighted portfolio | NYSE and AMEX stocks (1968 – 1998) | Sharpe ratio |
| Polson and Tew (2000) | Found significant out-performance of index using daily data for volatility estimation and upper and lower bounds on asset weights | S&P 500 Daily (1970 - 1996) | Distance to the ex-post efficient frontier of optimised portfolio relative to the S&P index |
| Wang (2005) | Aversion to model uncertainty leads to portfolios that are not mean-variance efficient | NYSE, AMEX and NASDAQ stocks (1963 - 1998) | Single period mean-variance utility function |
| Disatnik and Benninga (2007) | Found no advantage in using shrinkage estimation of covariance matrix. Averages of sample covariance matrix with single index matrix and others worked just as well. | NYSE (1964 - 2003) | Ex-post standard deviations of returns of minimum variance portfolio |
| DeMiguel, Garlappi, Nogales and Uppal (2009) | Constraining the norm of the weight vector often leads to higher Sharpe ratios than shrinkage or other approaches | 4 sets of US portfolios (1963 - 2004) plus a CRSP set (1968 - 2005) | Out-of-sample portfolio variance, Sharpe ratio, and portfolio turnover. |
| Huang et al (2018) | Naive diversification outperforms optimal mean–variance diversification but tail risk increases with number of stocks. | CRSP stocks on NYSE, AMEX, NASDAQ, and NYSE Arca. (1963 - 2014) | VaR, expected shortfall and other tail risk criteria. |
| Yu et al. (2019) | Worst case Omega ratio outperformed CVaR strategies and equally weighted portfolio | ETFs on MSCI indices for 21 countries (2001 - 2012) and Composition stocks from S&P 500 (2003 - 2014) | Realised portfolio value. |

Table 1 summarises the findings of some empirical comparisons of portfolio selection methods. Looking at the conclusion column in Table 1, we see little support for using vanilla mean-variance optimisation. There are mixed conclusions about naïve diversification, but there is support for schemes constraining the vector of asset weights. Shrinkage estimation of mean returns receives some backing as does shrinkage estimation of the covariance matrix, but simpler ways of averaging estimates are reported to work as well. More recently introduced objective functions such as the Omega ratio appear worthy of further empirical investigation.

### 3. Extensions of the Berkowitz test

A formal testing procedure is necessary to judge if ex-post portfolio returns are consistent with ex-ante estimates. Our procedure uses density forecasting, the prediction of the density function of a random variable. Here the random variable is the out-of-sample return on a portfolio selected at time $t$, $R_t$. This return is forecast as having a cumulative density function (*cdf*), $F_t(\ )$. Since $F_t(\ )$ may change over time, the approach commonly followed in measuring the accuracy of the density forecast is to use the empirical *cdf*. The measurement of the accuracy of the density forecast is then based on a sequence of observations, $r_t$, of realised out-of-sample returns and corresponding predicted *cdf*s, $F_t(r_t)$. One approach to the comparison of density functions develops the use of Kolmogorov-Smirnov methodology for density comparison. Recent developments and the background to this approach are given by Corradi and Swanson (2006). Another approach centres on the uniform distribution: Diebold, Gunther and Tay (1998) note that if the *cdf* is predicted accurately, then it should be a uniform random variable between zero and one; i.e. $F_t(r_t) \sim U(0,1)$. Hong and Li (2005) and Hong, Li and Zhao (2007) develop this approach.

Mitchell and Hall (2005) and Bao, Lee and Saltoglu (2007) give full discussions of the measurement of density forecasting accuracy. In line with their findings, we use an approach due to Berkowitz (2001) who reasoned that since statistical tests are well developed to detect departures from normality, a normal transformation should be used in this context. Firstly, the null hypothesis is that the *cdf* of the out-of-sample return, $F_t(r_t)$, has been correctly identified. Secondly, the null hypothesis then becomes that the sequence of observed values of $F_t(r_t)$, $\hat{F}_t(r_t)$, should be $U(0,1)$ distributed. Berkowitz develops the null hypothesis to a third stage, where the *cdf*, $F_t(r_t)$, is transformed to a standardised normal random variable, $y_t = \Phi^{-1}\left(\hat{F}_t(r_t)\right)$. We consider the case where

$$y_t - \mu = \varepsilon_t, \tag{1}$$

where $V(\varepsilon_t) = \sigma^2$. For $K$ observed returns $r_k$ $k=1,...K$, the log-likelihood function is

$$L(\mu,\sigma^2) = -\frac{K}{2}\ln(2\pi\sigma^2) - \sum_{k=1}^{K}\left(\frac{(y_k-\mu)^2}{2\sigma^2}\right). \tag{2}$$

The likelihood ratio test statistic is

$$\text{Berkowitz statistic} = -2\left(L(0,1) - L(\hat{\mu},\hat{\sigma}^2)\right), \tag{3}$$

where $\hat{\mu}, \hat{\sigma}$ are the sample mean and standard deviation of the $y_k$. Under the null hypothesis that $F(.)$ is correctly identified, this places two restrictions on the likelihood function and thus asymptotically the Berkowitz statistic is distributed as $\chi_2^2$.

However, in the analysis reported in this paper, the derivation of $\hat{F}_t(r_t)$ is based on a rolling finite sample of size *M*. Thus we investigate whether some adjustment is necessary for the use of overlapping observations and for small values of *M* or *K*. The experiment carried out and the resulting critical values developed are described in Appendix A.

Further, we consider the case where the process underlying the generation of $R_t$, and consequently of $y_t$, is subject to change. The Berkowitz statistic uses an equally weighted moving average of likelihood terms, see (2). To make the test more sensitive to recent observations of $y_t$, an obvious next step is to use an exponentially weighted moving average (*ewma*) that will react to a change in in the distribution of $R_t$ as it occurs but will taper off its importance as time passes. To implement this innovation, we revise (2) such that for *K* observed returns $r_k$ *k=1,...K*, the log-likelihood function is

$$L(\mu,\sigma^2) = \frac{-\sum_{k=1}^{K}\gamma^{K-k}\left(\frac{1}{2}\ln(2\pi\sigma^2) + \frac{(y_k-\mu)^2}{2\sigma^2}\right)}{\sum_{k=1}^{K}\gamma^{K-k}}, \tag{4}$$

where $0 < \gamma \leq 1$. The critical values for each value of the discount factor, $\gamma$, are generated using the same process described in Appendix A.

## 4. Construction of the consistency frontier

In order to discover the region in risk-expected return space where a selected portfolio produces out-of-sample returns that are consistent with the in-sample portfolio mean and variance, we explore the space dominated by the efficient frontier. The methodology used for the identification of consistency regions is independent of that used to establish the efficient frontier. Thus, for simplicity, we use mean-variance analysis to determine the efficient frontier. In addition, we consider long only portfolios to avoid the issues associated with short selling such as excessive transaction costs and its possible prohibition in some countries.

To define the position of portfolios in risk-expected return space, we set up a grid bounded by the efficient frontier. For each time origin considered, the scale on the expected return axis of the grid runs from the expected return of the minimum variance portfolio through to the maximum expected return portfolio. The scale on the risk axis is conditional on the level of expected return and is defined by the range the volatility of a portfolio can take. For each point on this grid, the sequence of in-sample densities and the corresponding sequence of out-of-sample returns allows the construction of a time series of realised probabilities; i.e. the probability of the portfolio achieving the observed out-of-sample return, given its in-sample density function. These probabilities allow the accuracy of the in-sample densities to be tested using the Berkowitz statistic. Example plots of the efficient frontier and the grid of points in risk-expected return space are shown below in Figure 1. At each grid point, a portfolio is deemed consistent when the Berkowitz statistic indicates that the ex-post return density is consistent with the ex-ante parameters.

We first describe the construction of the efficient frontier and then its role in the grid in risk-expected return space. The computation of the Berkowitz statistic and the determination of the consistency region are then described.

*4.1 Construction of the efficient frontier*

At a given time period, $t_k$, we construct an efficient frontier of portfolios from a universe of $N$ assets using an estimation interval of $M$ historical observations of returns. The relevant part of this frontier runs from the minimum variance portfolio to the portfolio with the highest expected return. The weights on the assets in the portfolio are denoted by the column vector $\boldsymbol{\omega}$, its transpose by $\boldsymbol{\omega}'$, the weights $\boldsymbol{\omega}_0$ in the minimum variance portfolio are the solution to

$$\textit{Minimise } \boldsymbol{\omega}'\hat{\boldsymbol{\Sigma}}_{t_k}\boldsymbol{\omega} \quad \textit{subject to } 0 \leq \omega_i \leq U \ \forall_i \textit{ and } \sum_{i=1}^{N}\omega_i = 1, \qquad (5)$$

where $U$ is an upper bound on the weight of any individual asset. The expected return and risk of on the minimum variance portfolio are, respectively

$$R_0(t_k) = \sum_{i=1}^{N}\omega_{0,i}\hat{\mu}_{i,t_k} \ ; \ SD_0(t_k) = \sqrt{\boldsymbol{\omega}_0'\hat{\boldsymbol{\Sigma}}_{t_k}\boldsymbol{\omega}_0}.$$

The ^ symbol denotes that the covariance matrix and expected return vector are estimated from sample data.

*4.2 Construction of a grid*

A crucial step in the method presented in this paper is the definition of a grid that identifies the position of a portfolio in risk-expected return space. This structure allows the properties of portfolios in a given grid position to be analysed over time. The risk axis of the grid is defined by evaluating the efficient frontier at $B$ equally spaced points (on the return axis), $(R_0(t_k), R_1(t_k), \ldots, R_{B-1}(t_k))$ where $R_B(t_k)$ is the expected return of the maximum expected return portfolio determined subject to the constraints in (5).

To identify the dominated portfolios within the efficient frontier, *RP* random portfolios are generated for each of these expected return values, satisfying both the constraints in (5) plus $R_b(t_k) = \sum_{i=1}^{N} \omega_i \hat{\mu}_{i,t_k}$. The range of standard deviations of these portfolios with expected return $R_b(t_k)$ is defined by *C* equally spaced points on the standard deviation axis, $(SD_{b,0}(t_k), SD_{b,1}(t_k), \ldots, SD_{b,c}(t_k), \ldots, SD_{b,C-1}(t_k))$, where $SD_{b,C-1}(t_k)$ is the maximum portfolio standard deviation found among the random portfolios. To summarise, at time $t_k$, we divide the region inside the efficient frontier into a grid with *B×C* points, where a point (*b,c*) in the grid has expected return $R_b(t_k)$ and standard deviation $SD_{b,c}(t_k)$. In the results reported below *RP* is set to 500.

Inside the efficient frontier, specification of the mean and standard deviation for a portfolio at grid point (*b,c*) does not define the portfolio weights unambiguously. To achieve as smooth a transition (in weights and in subsequent forecasting behaviour) as possible between portfolios in neighbouring grid positions, we define the portfolio weights by minimising their variance. Hence the weights are the solution to

$$Min(Variance(\omega_1, \ldots, \omega_N)) \text{ subject to: } 0 \le \omega_i \le U \, \forall i \text{ and } \sum_{i=1}^{N} \omega_i = 1,$$

$$\sum_{i=1}^{N} \omega_i \hat{\mu}_{i,t_k} = R_b(t_k) \text{ and } \boldsymbol{\omega}' \hat{\boldsymbol{\Sigma}}_{t_k} \boldsymbol{\omega} = \left( SD_{b,c}(t_k) \right)^2. \tag{6}$$

This minimisation with a non-linear constraint uses an algorithm due to Spellucci (1998a, 1998b); the starting point for the minimisation is the nearest random portfolio.

### *4.3 Computation of the Berkowitz statistic*

The Berkowitz statistic considers the out-of-sample performance of a portfolio given its historical performance. We compute the realised value of $r_{t_k,H}^{b,c}$, the out-of-sample portfolio return over an investment horizon of *H* time periods from $t_k+1$ to $t_k+H$ for each of the *B×C* portfolios considered. We hypothesise that the out-of-sample portfolio return is drawn from a distribution defined by its historical (in-sample) performance. To test this hypothesis, the empirical *cdf* of the out-of-sample return is computed using the in-sample returns. There are *M–H*+1 in-sample (overlapping) returns: $\left( r_{t,H}^{b,c} \mid \Theta_t \text{ for } t = t_k - M + 1 \text{ to } t_k - H + 1 \right)$ where $\Theta_t$ represents the information set available at time $t$. In this derivation of the empirical *cdf* of the out-of-sample return, $\hat{F}(R_{t_k,H}^{b,c} | \Theta_{t_k})$, we will drop the superscripts denoting grid position.

Under the null hypothesis, the out-of-sample portfolio return is drawn from a distribution with *cdf* $F(R_{t_k,H})$ which is estimated from its historical (in-sample) returns. There are $n = M-H+1$ in-sample (overlapping) returns: $(r_{t,H} | \Theta_{t_k}$ for $t = t_k - M + 1$ to $t_k - H + 1)$. Further simplifying our notation, we consider a set of returns, $r_1, \ldots, r_n$, which form the in-sample data; the out-of-sample return is $r_{n+1}$. From the in-sample data we define $r'_L$ as the $L^{th}$ observed return in ascending order of magnitude from $r'_1 = \min(r_1, \ldots, r_n)$ to $r'_n = \max(r_1, \ldots, r_n)$. We also define $\hat{\mu}_R$ and $\hat{\sigma}_R$ as the sample mean and standard deviation of $r_1, \ldots, r_n$. For the calculation of the empirical *cdf*, we consider two cases. The first case is when the out-of-sample return falls within the in-sample range, $r'_1 < r_{n+1} < r'_n$

$$\hat{F}(r_{n+1} | \Theta) = \frac{L + \frac{1}{2} + \left(\frac{r_{n+1} - r'_L}{r'_{L+1} - r'_L}\right)}{n+1}, \tag{7}$$

where there are $L$ in-sample observations less than $r_{n+1}$ and $r'_L$ is the greatest in-sample observation less than $r_{n+1}$. In the second case, $r_{n+1}$ falls outside the in-sample range, $r_{n+1} \leq r'_1$ or $r_{n+1} \geq r'_n$ and we use the normal distribution to approximate the behaviour of the out-of-sample return

$$\hat{F}(r_{n+1} | \Theta) = \begin{cases} \dfrac{1}{2(n+1)} \left( \dfrac{\Phi\left(\dfrac{r_{n+1} - \hat{\mu}_R}{\hat{\sigma}_R}\right)}{\Phi\left(\dfrac{r'_1 - \hat{\mu}_R}{\hat{\sigma}_R}\right)} \right) & \text{if } r_{n+1} \leq r'_1 \\[2em] 1 - \dfrac{1}{2(n+1)} \left( \dfrac{1 - \Phi\left(\dfrac{r_{n+1} - \hat{\mu}_R}{\hat{\sigma}_R}\right)}{1 - \Phi\left(\dfrac{r'_n - \hat{\mu}_R}{\hat{\sigma}_R}\right)} \right) & \text{if } r_{n+1} \geq r'_n \end{cases} \tag{8}$$

Under the null hypothesis the observed distribution of out-of-sample returns is consistent with the observed in-sample return distribution; i.e. the null hypothesis is

$$\text{the } cdf \text{ of the random variable } r_{t_k,H}^{b,c} \text{ is } \hat{F}(R_{t_k,H}^{b,c} | \Theta_{t_k}). \tag{9}$$

*4.4 Establishing consistency*

To measure the accuracy of the density forecast provided by the empirical *cdf*, we use repeated observations for each grid point (*b,c*). We consider portfolio selection at *K* consecutive time origins, $(t_1, t_2, \ldots, t_k, \ldots, t_K)$. At each origin $t_k$, for $k = 1, \ldots, K$, for each portfolio with expected return $R_b(t_k)$ and standard deviation, $SD_{b,c}(t_k)$ the within-sample historical portfolio returns and the out-of-sample return are calculated and the empirical *cdf* of the out-of-sample return computed. For each grid point (*b,c*), we have a series of *K* empirical *cdf*s which is used to compute the Berkowitz statistic.

To discover the values of (*b,c*) where ex-post portfolio performance is consistent with ex-ante estimates, we are concerned with ***accepting*** the null hypothesis (9), rather than with rejecting it as is more common in hypothesis testing. In order to make acceptance of the null hypothesis a demanding requirement, we set the probability of rejecting the null when it is true to a higher value, than would normally be used if the concern was the standard detection of departures from the null. In the results reported below the value of 20% is used.

## 5. The ex-post mathematics of efficient portfolios

As noted in Section 2, the empirical evidence suggests that the ex-post performance of an efficient portfolio is often inferior to that predicted ex-ante. The ex-post properties of idealised efficient portfolios, constructed subject only to the restriction of the budget constraint, are reported in Adcock (2013). These results, which are summarised and extended in this section, provide theoretical support for the methods and results reported in this paper. The notation, descriptions and text in Sections 5.1 and 5.2 below follow Adcock (2013). Section 5.3 presents extended results for the case of restrictions under a number of linear equality constraint. In general, efficient portfolios are constructed subject to inequality constraints, such as the requirement that portfolio weights be non-negative. A consequence of such inequality constraints is that the mathematics of the efficient frontier is intractable. Accordingly, Section 5.4 describes a simulation method which allows estimation of the ex-post efficient frontier in the presence of inequality constraints.

### *5.1 Traditional efficient set mathematics*

Let **R** be an *n*-vector of asset returns, which has the multivariate normal distribution $N(\boldsymbol{\mu}, \boldsymbol{\Sigma})$. The notation $R_p$ denotes portfolio return and $r_f$ the risk free rate. The notations ***1***, $\mathbf{0}_n$ and $\mathbf{0}_{mn}$ denote respectively an *n*-vector of ones, an *n*-vector of zeros and an $m \times n$ matrix of zeros. Subscripts are generally omitted. It is assumed that the covariance matrix $\boldsymbol{\Sigma}$ is non-singular. Maximising expected utility subject only to the budget constraint and recalling Stein's Lemma, the first order conditions for portfolio selection lead to the well-known expression for portfolio weights

$$w = \frac{\boldsymbol{\Sigma}^{-1} \mathbf{1}}{\mathbf{1}^T \boldsymbol{\Sigma}^{-1} \mathbf{1}} + \theta \left\{ \boldsymbol{\Sigma}^{-1} - \frac{\boldsymbol{\Sigma}^{-1} \mathbf{1} \mathbf{1}^T \boldsymbol{\Sigma}^{-1}}{\mathbf{1}^T \boldsymbol{\Sigma}^{-1} \mathbf{1}} \right\} \boldsymbol{\mu} = w_0 + \theta w_1; \theta \geq 0.$$

The vector $w_0$ is the minimum variance portfolio and satisfies the budget constraint $\mathbf{1}^T w_0 = 1$. The vector $w_1$ is a self-financing portfolio[2]. Risk appetite $\theta$ is defined as

$$\theta = -E\{U'(R_p)\}/E\{U''(R_p)\}.$$

The expected return and variance of portfolio return, which has a normal distribution given the assumption that $\mu$ and $\Sigma$ are given, are

$$\mu_p = \alpha_0 + \theta\alpha_1, \; \sigma_p^2 = \alpha_2 + \theta^2\alpha_1,$$

respectively, where the standard constants are defined as

$$\alpha_0 = \frac{\mu^T \Sigma^{-1} \mathbf{1}}{\mathbf{1}^T \Sigma^{-1} \mathbf{1}}, \alpha_1 = \mu^T \left\{ \Sigma^{-1} - \frac{\Sigma^{-1} \mathbf{1} \mathbf{1}^T \Sigma^{-1}}{\mathbf{1}^T \Sigma^{-1} \mathbf{1}} \right\} \mu, \; \alpha_2 = \frac{1}{\mathbf{1}^T \Sigma^{-1} \mathbf{1}}.$$

Note that these definitions of the standard constants are the same as those used in Kan and Smith (2008). The equation of the efficient frontier is

$$\mu_p - \alpha_0 = \sqrt{\alpha_1}\sqrt{\sigma_p^2 - \alpha_2}.$$

*5.2 Ex-post distribution of portfolio returns*

When $\mu$ is replaced by a forecast, denoted $F$, portfolio return is distributed as an extended quadratic form in normal variables. It is assumed the *2n*-vector

$$X = \begin{bmatrix} R \\ F \end{bmatrix},$$

has a non-singular multivariate normal distribution $N(\tau, \Gamma)$ with

$$\tau = \begin{bmatrix} \mu \\ \mu + \delta \end{bmatrix}, \Gamma = \begin{bmatrix} \Sigma_{RR} & \Sigma_{RF} \\ \Sigma_{FR} & \Sigma_{FF} \end{bmatrix},$$

respectively. Non-zero entries in the vector $\delta$ mean that the forecast is biased. It is assumed that the covariance matrix is known. The vector of portfolio weights based upon the forecast $F$ is

$$w = w_0 + \theta \mathbf{D}_0 F, \; \mathbf{D}_0 = \left\{ \Sigma^{-1} - \frac{\Sigma^{-1} \mathbf{1} \mathbf{1}^T \Sigma^{-1}}{\mathbf{1}^T \Sigma^{-1} \mathbf{1}} \right\}.$$

Portfolio return is

$$R_p = a^T X + 2^{-1} X^T \mathbf{A} X,$$

with

$$a = \begin{bmatrix} w_0 \\ \mathbf{0}_n \end{bmatrix}, \mathbf{A} = \begin{bmatrix} \mathbf{0}_{nn} & \theta \mathbf{D}_0 \\ \theta \mathbf{D}_0 & \mathbf{0}_{nn} \end{bmatrix},$$

---

[2] In this section vectors of portfolio weights are denoted using w; the notation $\omega$ is reserved for portfolios designed using the methods reported in Section 4.

which is distributed as an extended quadratic form in normal variables. The properties of these are described in detail in Mathai and Prevost (1992). Relevant results for financial applications are in Appendix B of Adcock et al (2012). The probability density function of $R_p$ is intractable, although the central limit theorem means that, *ceteris paribus,* its distribution will tend to normality as the number of assets increases. The characteristic function, however, is tractable and leads to analytical results for the mean and variance of portfolio returns. Proposition 2 of Adcock (2013) proposition is as follows. The expected value and variance of portfolio return, denoted by the additional subscript *f*, are respectively

$$\mu_{pf} = \alpha_0 + \theta\alpha_1 + \theta\beta_0, \; \sigma^2_{pf} = \alpha_2 + \theta^2\alpha_1 + 2\theta\beta_1 + \theta^2\beta_2,$$

where $\beta_0$ and $\beta_1$ are

$$\beta_0 = trace\{\mathbf{D}_0\mathbf{\Sigma}_{RF}\} + \boldsymbol{\mu}^T\mathbf{D}_0\boldsymbol{\delta}, \; \beta_1 = \boldsymbol{\mu}^T\mathbf{D}_0\mathbf{\Sigma}_{FR}\mathbf{w}_0,$$

and

$$\beta_2 = trace\{(\mathbf{D}_0\mathbf{\Sigma}_{RF})^2 + \mathbf{D}_0\mathbf{\Sigma}_{FF}\mathbf{D}_0\mathbf{\Sigma}_{RR}\} + \boldsymbol{\mu}^T\mathbf{D}_0\mathbf{\Sigma}_{FF}\mathbf{D}_0\boldsymbol{\mu} + \boldsymbol{\delta}^T\mathbf{D}_0\mathbf{\Sigma}_{RR}\mathbf{D}_0\boldsymbol{\delta}$$

$$+ 2\boldsymbol{\delta}^T\mathbf{D}_0\mathbf{\Sigma}_{RR}\mathbf{D}_0\boldsymbol{\mu} + 2\boldsymbol{\mu}^T\mathbf{D}_0\mathbf{\Sigma}_{FR}\mathbf{D}_0(\boldsymbol{\mu} + \boldsymbol{\delta}).$$

Substitution gives the following corollary from (Adcock, 2013). The equation of the efficient frontier is

$$\mu_{pf} = A_0 + \left(A_1/\sqrt{B_1}\right)\sqrt{\sigma^2_{pf} - B_0},$$

where

$$A_0 = \alpha_0 - \beta_1(\alpha_1 + \beta_0)/(\alpha_1 + \beta_2), \; A_1 = (\alpha_1 + \beta_0),$$

$$B_0 = \alpha_2 - \beta_1^2/(\alpha_1 + \beta_2), \; B_1 = (\alpha_1 + \beta_2).$$

Thus, the *ex-post* expected return and variance of an efficient portfolio constructed using estimates or forecasts of expected returns are different from those based on standard efficient set mathematics. As an exemplar, consider the case where the bias $\boldsymbol{\delta}$ is a vector of zeros and the covariance matrix $\boldsymbol{\Gamma}$ is

$$\boldsymbol{\Gamma} = \begin{bmatrix} \boldsymbol{\Sigma}_{RR} & \rho\sqrt{\kappa}\boldsymbol{\Sigma}_{RR} \\ \rho\sqrt{\kappa}\boldsymbol{\Sigma}_{RR} & \kappa\boldsymbol{\Sigma}_{RR} \end{bmatrix},$$

where $\kappa > 0$ is a specified constant. If it is further assumed that the correlation between actual return and forecast for asset *i* are equal and that cross correlations are zero the constants $\beta_{0,1,2}$ are

$$\beta_0 = \rho(n-1)\sqrt{\kappa}, \; \beta_1 = 0, \; \beta_2 = \kappa\{\alpha_1 + (n-1)(1+\rho^2)\} + 2\rho\alpha_1\sqrt{\kappa},$$

leading to

$$\mu_{pf} = \alpha_0 + \theta\{\alpha_1 + \rho(n-1)\sqrt{\kappa}\},\ \sigma^2_{pf} = \alpha_2 + \theta^2\{(1+\kappa)\alpha_1 + (n-1)(1+\rho^2) + 2\rho\alpha_1\sqrt{\kappa}\}.$$

A special case is the use of the sample mean returns of an IID time series of length $M$. In this case $\kappa = M^{-1}$ and $\rho = 0$, leading to

$$\mu_{pf} = \alpha_0 + \theta\alpha_1 = \mu_p,\ \sigma^2_{pf} = \alpha_2 + \theta^2\{(1+M^{-1})\alpha_1 + (n-1)\} = \sigma^2_p + \theta^2\{(n-1) + \alpha_1/M\}.$$

Thus, there is no effect on mean return, but there is an increase in variance. In particular, the variance is an increasing function of the number of assets. The value $\kappa = 1$ may be interpreted as meaning that the covariance matrix associated with the forecasts is predictive, which corresponds with sensible practice and which leads to

$$\sigma^2_{pf} = \alpha_2 + \theta^2\{2\alpha_1 + (n-1)\} = \sigma^2_p + \theta^2\{(n-1) + \alpha_1\}.$$

### 5.3 A Modification for several equality constraints

In practice, portfolio selection is carried out subject to inequality constraints, conventionally written as $\mathbf{A}^T w \geq \mathbf{b}$. Denoting the constraints that are active at the solution using the subscript $a$, the vector of portfolio weights based upon the forecast $F$ is

$$w = \widetilde{w}_0 + \theta \widetilde{\mathbf{D}}_0 F,$$

where

$$\widetilde{w}_0 = \Sigma^{-1}\mathbf{A}_a(\mathbf{A}_a^T\Sigma^{-1}\mathbf{A}_a)^{-1}\mathbf{b}_a,\ \widetilde{\mathbf{D}}_0 = \{\Sigma^{-1} - \Sigma^{-1}\mathbf{A}_a(\mathbf{A}_a^T\Sigma^{-1}\mathbf{A}_a)^{-1}\mathbf{A}_a^T\Sigma^{-1}\}$$

In this case, the standard constants are

$$\widetilde{\alpha}_0 = \mu^T\Sigma^{-1}\mathbf{A}_a(\mathbf{A}_a^T\Sigma^{-1}\mathbf{A}_a)^{-1}\mathbf{b}_a,\ \widetilde{\alpha}_1 = \mu^T\{\Sigma^{-1} - \Sigma^{-1}\mathbf{A}_a(\mathbf{A}_a^T\Sigma^{-1}\mathbf{A}_a)^{-1}\mathbf{A}_a^T\Sigma^{-1}\}\mu,$$

$$\widetilde{\alpha}_2 = \mathbf{b}_a^T(\mathbf{A}_a^T\Sigma^{-1}\mathbf{A}_a)^{-1}\mathbf{b}_a.$$

The standard ex-ante mean and variance are

$$\mu_p = \widetilde{\alpha}_0 + \theta\widetilde{\alpha}_1,\ \sigma^2_p = \widetilde{\alpha}_2 + \theta^2\widetilde{\alpha}_1. \tag{10}$$

For the case above, in which the forecast is the sample mean returns of an IID time series, if there are $p$ active constraints, the *ex-post* mean is unchanged, but the portfolio variance is

$$\widetilde{\sigma}^2_{pf} = \widetilde{\alpha}_2 + \theta^2\{(1+T^{-1})\widetilde{\alpha}_1 + (n-p)\} = \widetilde{\sigma}^2_p + \theta^2\{(n-p) + \widetilde{\alpha}_1/M\},$$

or, if a predictive covariance matrix is used

$$\widetilde{\sigma}^2_{pf} = \widetilde{\alpha}_2 + \theta^2\{2\widetilde{\alpha}_1 + (n-p)\} = \widetilde{\sigma}^2_p + \theta^2\{(n-p) + \widetilde{\alpha}_1\}. \tag{11}$$

### 5.4 Estimating the ex-post efficient frontier in the presence of inequality constraints

As stated above, a consequence of inequality constraints is that the mathematics of the efficient frontier is intractable. The moments of the distribution of ex-post portfolio returns may, however, be estimated using simulation methods. For the model validation described in Section 6, it is assumed that the *2n*-vector time series of asset and forecast returns is IID multivariate normal as specified in (10). For a given mean-variance efficient portfolio the simulation procedure is as follows. For a specified sample size *P*, generate simulated forecasts $F_i; i = 1, \cdots, P$ from the multivariate normal distribution $N(\mu + \delta, \Sigma_{FF})$. For each of the *P* simulated vectors of forecasts compute the corresponding mean-variance efficient portfolio. This results in (a) vectors of portfolio weights $w_i; i = 1, \cdots, P$ and (b) matrices $A_{a,i}$ and vectors $b_{a,i}$ corresponding to the active set at each simulation. For convenience, we label the method described in Section 5.3 as Method 0.

Method 1 is non-parametric. Denoting the sample mean vector and sample covariance matrix of the set $\{w_i\}$ by $\phi$ and $\Omega$ respectively, and noting that for the forecasts used in model validation the cross covariance matrix $\Sigma_{RF}$ is a zero matrix and $\delta$ is a zero vector, the vector of estimated ex-post expected returns and estimated ex-post covariance matrix are, respectively

$$E(R_p^{EP}) = \phi^T \mu; cov(R_p^{EP}) = \phi^T \Sigma_{RR} \phi + \mu^T \Omega \mu + trace(\Sigma_{RR}\Omega).$$

This method may be employed if the joint distribution of returns and forecasts is time varying, but is multivariate normal as in (10) at time *t*. It may also be used if the joint distribution of returns and forecasts is not multivariate normal but mean-variance portfolio selection is performed. The method requires extensions to the scope of the simulations in the presence of non-zero cross-correlations.

## 6. Model validation

In this section, we demonstrate the computation of the consistency region and the consistency frontier. A validation of the ex-post efficient frontier and its consistency follows. Finally, we compare and contrast the consistency frontier and the ex-post efficient frontier.

### *6.1 Validation of the consistency region*

A set of multivariate normally distributed asset returns is generated using the estimated means and covariance matrix of the DJ30 constituent stocks, details of the data are given in Section 7.1. The time series of vectors of asset returns are generated following a Cholesky factorization of the covariance matrix. This simulated data set is used to produce the example efficient frontiers and dominated portfolios shown in Figure 1 and has *N* = 30 assets, with an upper bound on a holding of a single asset of *U* = 0.33. The six plots shown correspond to a range of values for *M* historical observations of returns, *M* = 52, 104, 156, 208, 260, 312 equivalent to estimation intervals of one to six years. The investment period *H* is chosen to be 4 weeks, the empirical *cdf* is found from the *M-H*+1 overlapping observations of 4 week returns. The sample mean vector and covariance matrix are used to estimate both the ex-ante efficient frontiers (using (5)) and the ex-post efficient frontier (using (10) and (11)). There are *B* = 11 values of expected portfolio return and *C* = 50 portfolios. Although the choice of portfolio standard deviations is designed to be equally spaced along the axis, slight irregularities occur for computational reasons, if a solution to (6) is not found (which may happen as (6) is a quadratically constrained quadratic program that is hard to solve numerically), the nearest random portfolio is used instead. To calculate the Berkowitz statistic, we consider $K = 39$ origins (since the investment horizon is four weeks, this equates to considering density forecasts over three years). The means and standard deviations of portfolio returns are averaged over the $K$ origins to determine the position in risk-expected return space of each point ($b,c$) in Figure 1. We present the values of the Berkowitz statistic (see (3)) for each point graphically. A (green) cross indicates where the null hypothesis (that the out-of-sample returns are generated by the observed in-sample distribution) is not rejected at 20%; if the hypothesis is rejected then a (red) dash is shown.

**Figure 1. Ex-ante efficient frontiers, ex-post efficient frontiers and dominated portfolios averaged over *K*=39 origins for a range of *M* estimation regions using a multivariate normal data set.** *The symbols indicate the significance of the Berkowitz statistic, a (red) circle indicates that the consistency hypothesis can be rejected at 20% significance, a (green) cross indicates acceptance.*

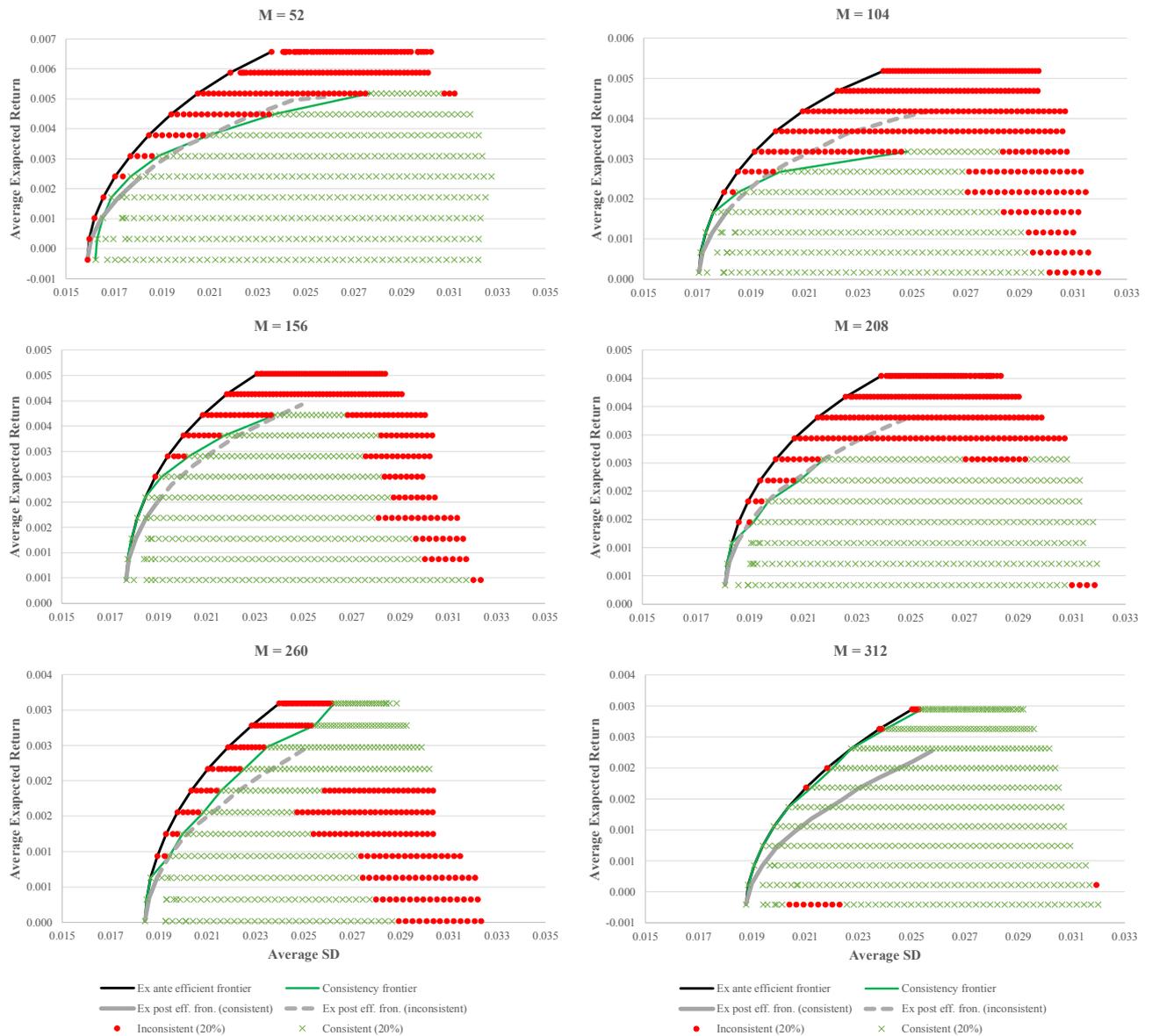

The (green) area of the graph where the null hypothesis is accepted is called the consistency region. Due to the stochastic nature of hypothesis testing where the value of the Berkowitz statistic is close to its critical value, the boundary of the consistent region is not smooth. To make the transition from rejection of the hypothesis to acceptance appear more smoothly for each band (of constant expected return) in the plots, we use a symmetric moving average using up to 9 values of the Berkowitz statistics centred on the value of $c$ as the criterion for acceptance and identification of the consistency region. Considering Figure 1, we bear in mind that the plots use ex-ante axes and the consistency region shows where the ex-post and ex-ante performances of a portfolio coincide. We see that the consistency region of portfolios peels away from the efficient frontier as the expected return increases. In other words, the greater expected portfolio returns on the efficient frontier are only deliverable at higher risk than that implied by the frontier, if at all. As the estimation period, *M,* increases: the range of expected returns decreases; the size of the consistency region tends to increase; the consistent frontier moves closer to the efficient frontier. In this context of multivariate normal returns with no time dependence, not surprisingly the use of more data for estimation improves the accuracy of the density forecasts implicit in the efficient frontier and the dominated portfolios within it.

*6.2 Validation of the ex-post efficient frontier*

The expected ex-post efficient frontiers, using Method 1, are also shown in Figure 1. The increased standard deviation of portfolio returns from the ex-ante frontier follows from (11). Overall, the ex-post efficient frontier and the consistency frontier are close to each other. For the lower bands of expected returns (usually $b \leq 5$) the ex-post frontier shows a slightly higher standard deviation than the consistency frontier. For $M \leq 208$, the higher expected portfolio returns on the ex-post efficient frontier are outside the consistency region. However, for $M \geq 260$, the ex-post frontier falls inside the consistency region. The consistency of the portfolios on the ex-post frontier is investigated using the Berkowitz statistic with the null hypothesis that the random variable $r_{t_k,H}^{b,0}$ is $N\left(H\mu_{pf}, H\tilde{\sigma}_{pf}^2\right)$, where $H = 4$, $\mu_{pf} = R_b(t_k)$ and $\sigma_p = SD_{b,0}(t_k)$ is input to (11) to form $\tilde{\sigma}_{pf}^2$. From Figure 1, we see that, except for *M* = 312, the ex-post efficient frontier becomes inconsistent for higher expected returns (for $b \geq 5$ or 6 depending on *M*). As explained earlier, we have defined consistency by a *p*-value of the Berkowitz statistic of 0.20 (20%) or more. The averaged (over *M*) p-values of the Berkowitz statistic for the ex-post frontier from *b* = 1, …, 11 are: (0.48, 0.43, 0.36, 0.28, 0.21, 0.17, 0.12, 0.13, 0.12, 0.12, 0.14). These values indicate that the accuracy of the predicted ex-post frontier declines as the expected return increases. Inspection of the standardised returns under the null hypothesis shows that for higher levels of expected return, there is an increasing tendency to overestimate the mean return and underestimate the standard deviation of returns.

*6.3 A comparison of the consistency frontier and the ex-post efficient frontier*

To investigate the degree of agreement between the consistency frontier (CF) and the ex-post efficient frontier (EF), we proceed as follows. For each of the six sample sizes and for each of the K = 39 origins, an efficient frontier is constructed with the same expected value as a corresponding portfolio on the consistency frontier. The portfolios are constructed subject to the same constraints used to build consistency frontiers. For each EF portfolio, the ex-post mean and volatility are estimated using the simulation methods described in Section 5.4. For each of the six sample sizes, the results are averaged over the K=39 time origins.

**Table 2. Differences between the volatility of portfolios along the CF and the corresponding ex-post volatilities for Methods 0 and 1, % difference is shown in italics. (na signifies no consistent portfolio)**

|  | | \multicolumn{12}{c}{Sample Size} | | | | | | | | | | | |
|---|---|---|---|---|---|---|---|---|---|---|---|---|---|
| | b = | 52 | | 104 | | 156 | | 208 | | 260 | | 312 | |
| Method 0 | 1 | -0.0003 | *-2.1* | 0.0000 | *0.0* | 0.0000 | *0.0* | 0.0000 | *0.0* | 0.0000 | *0.0* | 0.0000 | *0.0* |
| | 2 | -0.0003 | *-1.6* | 0.0001 | *0.4* | 0.0001 | *0.3* | 0.0001 | *0.4* | 0.0001 | *0.5* | 0.0001 | *0.6* |
| | 3 | 0.0000 | *-0.3* | 0.0002 | *1.5* | 0.0002 | *1.2* | 0.0002 | *1.3* | 0.0003 | *1.6* | 0.0003 | *1.8* |
| | 4 | 0.0003 | *1.9* | 0.0005 | *3.1* | 0.0004 | *2.4* | -0.0001 | *-0.4* | 0.0001 | *0.4* | 0.0007 | *3.4* |
| | 5 | 0.0004 | *2.5* | 0.0004 | *2.0* | 0.0007 | *3.7* | 0.0000 | *0.0* | 0.0003 | *1.5* | 0.0011 | *5.5* |
| | 6 | 0.0004 | *2.2* | -0.0002 | *-0.9* | 0.0008 | *4.3* | -0.0003 | *-1.7* | 0.0004 | *1.8* | 0.0016 | *8.0* |
| | 7 | -0.0002 | *-1.2* | -0.0037 | *-14.8* | 0.0007 | *3.4* | -0.0001 | *-0.5* | 0.0008 | *3.8* | 0.0021 | *9.8* |
| | 8 | -0.0012 | *-5.2* | na | *na* | 0.0006 | *2.5* | na | *na* | 0.0013 | *5.6* | 0.0027 | *12.5* |
| | 9 | -0.0029 | *-10.2* | na | *na* | 0.0001 | *0.5* | na | *na* | 0.0020 | *8.5* | 0.0036 | *15.7* |
| | 10 | na | *na* | na | *na* | na | *na* | na | *na* | 0.0018 | *7.1* | 0.0042 | *17.6* |
| | 11 | na | *na* | na | *na* | na | *na* | na | *na* | 0.0031 | *11.9* | 0.0051 | *20.0* |
| Method 1 | 1 | -0.0003 | *-2.1* | 0.0000 | *0.0* | 0.0000 | *0.0* | 0.0000 | *0.0* | 0.0000 | *0.0* | 0.0000 | *0.0* |
| | 2 | -0.0002 | *-1.5* | 0.0001 | *0.5* | 0.0001 | *0.4* | 0.0001 | *0.5* | 0.0001 | *0.6* | 0.0001 | *0.6* |
| | 3 | -0.0001 | *-0.4* | 0.0003 | *1.7* | 0.0002 | *1.3* | 0.0002 | *1.4* | 0.0003 | *1.5* | 0.0003 | *1.8* |
| | 4 | 0.0003 | *1.8* | 0.0005 | *2.7* | 0.0004 | *2.3* | -0.0001 | *-0.7* | 0.0000 | *-0.1* | 0.0005 | *2.8* |
| | 5 | 0.0002 | *1.2* | 0.0002 | *0.9* | 0.0005 | *2.8* | -0.0002 | *-1.1* | 0.0001 | *0.3* | 0.0008 | *3.9* |
| | 6 | -0.0001 | *-0.4* | -0.0006 | *-3.1* | 0.0005 | *2.7* | -0.0007 | *-3.3* | -0.0001 | *-0.5* | 0.0010 | *5.1* |
| | 7 | -0.0010 | *-4.9* | -0.0045 | *-18.2* | 0.0001 | *0.4* | -0.0008 | *-3.5* | -0.0001 | *-0.5* | 0.0010 | *4.8* |
| | 8 | -0.0025 | *-10.8* | na | *na* | -0.0005 | *-2.2* | na | *na* | -0.0002 | *-1.1* | 0.0009 | *4.1* |
| | 9 | -0.0054 | *-19.2* | na | *na* | -0.0018 | *-7.3* | na | *na* | -0.0004 | *-1.6* | 0.0012 | *5.2* |
| | 10 | na | *na* | na | *na* | na | *na* | na | *na* | -0.0014 | *-5.5* | 0.0007 | *3.0* |
| | 11 | na | *na* | na | *na* | na | *na* | na | *na* | -0.0011 | *-4.1* | 0.0003 | *1.3* |

Table 2 displays the difference between the volatility of portfolios along the CF and the corresponding ex-post volatilities. The results for Method 0 are computed using the results in Section 5.3; assuming that at each point on the constructed EF the constraints are equalities. The results for Method 1 are based on P = 10 simulations[3]. The rows for each panel of the table correspond to the expected returns on the grid. The columns of the table correspond to sample sizes 52 through 312. The first panel of the table shows the differences when ex-post volatility is computed using the methods of Section 5.3, assuming equality constraints. Panel 2 show the results for simulation Method 1 described in Section 5.4. For the dataset and methods used in this paper, Method 1 generally gives smaller errors. For both methods, the errors increase with expected return; that is, as one moves up the CF and EF. Unreported results suggest that the increase in error is associated with a larger number of portfolio weights that are at the upper permitted limit of 0.33. A similar exercise where the efficient frontier is constructed with the same volatility as a corresponding portfolio on the consistency frontier produced qualitatively similar results: errors are larger for points on the EF and CF which have higher expected returns.

---

[3] Results based on P = 50 simulations lead to conclusions that are essentially the same. This implies that for this data set and portfolio design parameters a small number of simulations is adequate for practical purposes. We suggest that this result is unsurprising as for forecasts based on *N* observations the volatility of forecasts is of order $1/\sqrt{N}$ and thus the number of simulated solutions with distinct active sets will be small.

The methods used to construct the consistency frontier and Method 1 are empirical and based on simulation. Consequently, it is to be expected that there will be differences in the various frontiers as shown in Table 2. To investigate the impact of these differences, we compute daily Conditional Value at Risk (CVaR) at 1% for all portfolios on the CF and the two ex-post EFs. Table 3 presents the difference in CVaR between portfolios on the CF and the ex-post EFs. As an exemplar, the entry for sample size 52 and point 1 on the CF, is a difference of 3 basis points for both methods of computing the ex-post EF.

**Table 3. The difference in CVaR between portfolios on the consistent frontier and the ex-post efficient frontier from Methods 0 and 1 (the values shown are in basis points of returns).**

|     | Sample Size | | | | | | | | | | | |
| --- | --- | --- | --- | --- | --- | --- | --- | --- | --- | --- | --- | --- |
|     | 52 | | 104 | | 156 | | 208 | | 260 | | 312 | |
| b = | Mtd 0 | Mtd 1 | Mtd 0 | Mtd 1 | Mtd 0 | Mtd 1 | Mtd 0 | Mtd 1 | Mtd 0 | Mtd 1 | Mtd 0 | Mtd 1 |
| 1 | 3 | 3 | 0 | 0 | 0 | 0 | 0 | 0 | 0 | 0 | 0 | 0 |
| 2 | 3 | 2 | 1 | 1 | 1 | 1 | 1 | 1 | 1 | 1 | 1 | 1 |
| 3 | 0 | 1 | 3 | 3 | 2 | 2 | 2 | 2 | 3 | 3 | 3 | 3 |
| 4 | 3 | 3 | 5 | 5 | 4 | 4 | 1 | 1 | 1 | 0 | 7 | 6 |
| 5 | 4 | 2 | 4 | 2 | 7 | 5 | 0 | 2 | 3 | 1 | 11 | 8 |
| 6 | 4 | 0 | 2 | 6 | 8 | 5 | 3 | 7 | 4 | 1 | 16 | 11 |
| 7 | 2 | 9 | 37 | 45 | 7 | 1 | 1 | 7 | 8 | 1 | 21 | 11 |
| 8 | 12 | 25 | na | na | 6 | 4 | na | na | 13 | 2 | 28 | 10 |
| 9 | 29 | 53 | na | na | 1 | 17 | na | na | 20 | 3 | 36 | 12 |
| 10 | na | na | na | na | na | na | na | na | 18 | 13 | 43 | 8 |
| 11 | na | na | na | na | na | na | na | na | 32 | 10 | 51 | 4 |

Table 3 suggests the following conclusions. Method 0, in which the constraints are treated as equalities, is almost as accurate or as accurate as Method 1 for estimation periods of 208 weeks or less. For five and six years, Method 1 is superior. Method 0 can be used in practice for estimation periods of 4 years or less and for portfolios that correspond to points 1 to 7 on the consistency frontier; that is for lower levels of risk. This has the advantage of avoiding the need for the simulations that are required for Method 1. The difference in CVaR is larger for EF portfolios in which weights are more likely to be at their permitted upper limits. We suggest that this is a consequence of the greater sensitivity of portfolio weights to expected returns when a constraint is active or almost active. Overall, the two methods, based on the theoretical results reported in Section 5.2 and extended in 5.3, confirm the results based on the procedures that lead to the consistency frontier.

## 7. Empirical study of DJ30 stocks

Using the constituent stocks of the Dow Jones Industrial Average, we compute the consistency region from 1997 to mid-2015. After describing the data, we discuss the behaviour of the consistency region based on actual data and contrast it with its behaviour in Section 6. The investment strategy implications of the consistency region are explored using the extensions to the Berkowitz statistic described in Section 3.

### 7.1 Data

The data set runs from 4th January 1988 to 27 July 2015. We use the constituents of the Dow Jones Industrial Average as of March 2015, However since the prices of four of these companies were not available for the full period and we replaced them with four previous constituents which had belonged to the index up to 2009 or 2013. We denote this data set as DJ30. The companies used are listed in Table 4 and summary statistics of the index and its constituents are given in Table 5.

**Table 4. A list of the companies used in DJ30 data set with the year they entered or exited the index**

| Company | Entry Date | Exit Date | Company | Entry Date | Exit Date |
|---|---|---|---|---|---|
| 3M | 1976 | | IBM | 1979 | |
| Alcoa | 1959 | 2013 | Intel | 1999 | |
| American Express | 1982 | | Johnson & Johnson | 1997 | |
| Apple | 2015 | | JPMorgan Chase | 1991 | |
| AT&T | 1916 | 2015 | McDonald's | 1985 | |
| Bank of America | 2008 | 2013 | Merck | 1979 | |
| Boeing | 1987 | | Microsoft | 1999 | |
| Caterpillar | 1991 | | Nike | 2013 | |
| Chevron | 1930 | 1999 | Pfizer | 2004 | |
| | 2008 | | Procter & Gamble | 1932 | |
| Cisco Systems | 2009 | | The Home Depot | 1999 | |
| Coca-Cola | 1987 | | Travelers | 2009 | |
| DuPont | 1935 | | United Technologies | 1939 | |
| ExxonMobil | 1928 | | Verizon | 2004 | |
| Hewlett Packard | 1997 | 2013 | Wal-Mart | 1997 | |
| General Electric | 1907 | | Walt Disney | 1991 | |

The summary statistics for skewness and kurtosis of the weekly return data indicate departure from the Normal distribution. A partial explanation of this departure is apparent in Figure 2, where the time series of the index, along with annualised moving averages of the index mean return and its volatility reveal the extent to which these moments of index return vary over time.

**Table 5. Summary statistics of weekly returns on the DJ 30 Index and its constituents**

| | | DJ Constituents | | | | |
|---|---|---|---|---|---|---|
| Statistic | DJ 30 Index | Minimum | Lower Quartile | Median | Upper Quartile | Maximum |
| **Mean (% p.a.)** | 1.11 | 0.59 | 1.38 | 1.64 | 1.79 | 2.78 |
| **SD (% p.a.)** | 16.18 | 20.40 | 23.57 | 26.95 | 32.50 | 44.48 |
| **Skewness** | -0.81 | -2.97 | -0.51 | -0.33 | -0.12 | 0.12 |

| Kurtosis | 7.0 | 1.7 | 2.9 | 4.2 | 7.1 | 42.8 |

### 7.2. Consistency regions computed using DJ30 data

The behaviour of the Dow Jones index, summarised in Figure 2, exhibits time dependence in both mean and variance. We use the Ledoit and Wolf (2003) method to estimate the covariance matrix of returns. Following our findings in Section 6, summarised in Figure 1, that the effect of sampling error is lessened by increasing $M$, we use $M = 312$ (equivalent to six years data) to calculate the estimated expected return and standard deviation of the portfolios on, and dominated by, the efficient frontier. Using an investment horizon of $H = 4$ weeks and $K = 39$ (equivalent to three years), 243 realisations of the Berkowitz statistic are computed from 23/12/1996 to 20/7/2015. Using the same format as Figure 1, in Figure 3 we show plots for four dates selected roughly four years apart culminating at the latest available date 20/7/2015.

**Figure 2. The Dow Jones Industrials Index from January 1988 to July 2015 is shown on left hand axis, an annualised return on the index (dashed line) and the index volatility (solid line) are shown on right hand axis. The evaluation dates used in Figure 3 are indicated.**

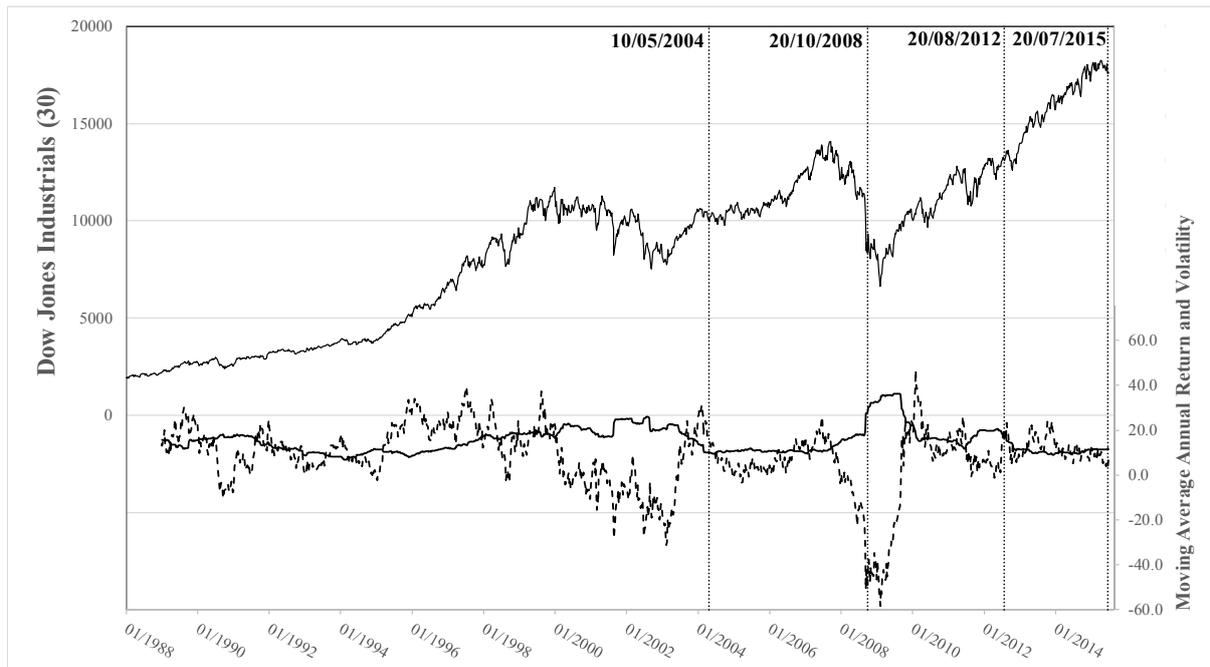

In Figure 1, with simulated multivariate normal data, for $M = 312$, the consistency region almost covers the entire set of dominated portfolios. As a contrast using actual data, we see in Figure 3 that the size of the consistency region fluctuates substantially over time. For 10/5/2004, the consistency region reaches $b = 7$; for 20/10/2008, the consistency region reaches $b = 3$; for 20/8/2012, the region returns to $b = 7$. For 20/7/2015, there are no consistent portfolios. The size of the consistency region is clearly time dependent. For the first three cases, investors may choose a portfolio on the consistency frontier and expect to receive out-of-sample performance defined by its ex-ante risk-expected return co-ordinates. The plot for 20/7/2015 demonstrates that at some points in time there is a lack of consistent portfolios. In these conditions, an investor choosing any portfolio, efficient or inefficient, would fail to receive the out-of-sample returns implied by the portfolio's in-sample performance and moving to cash would be a rational decision.

**Figure 3. Summary efficient frontiers for a universe of 30 members of the Dow Jones over *K*=39 origins for an estimation region of *M* = 312 for the four evaluation dates shown.** *(The symbols are as defined in Figure 1.)*

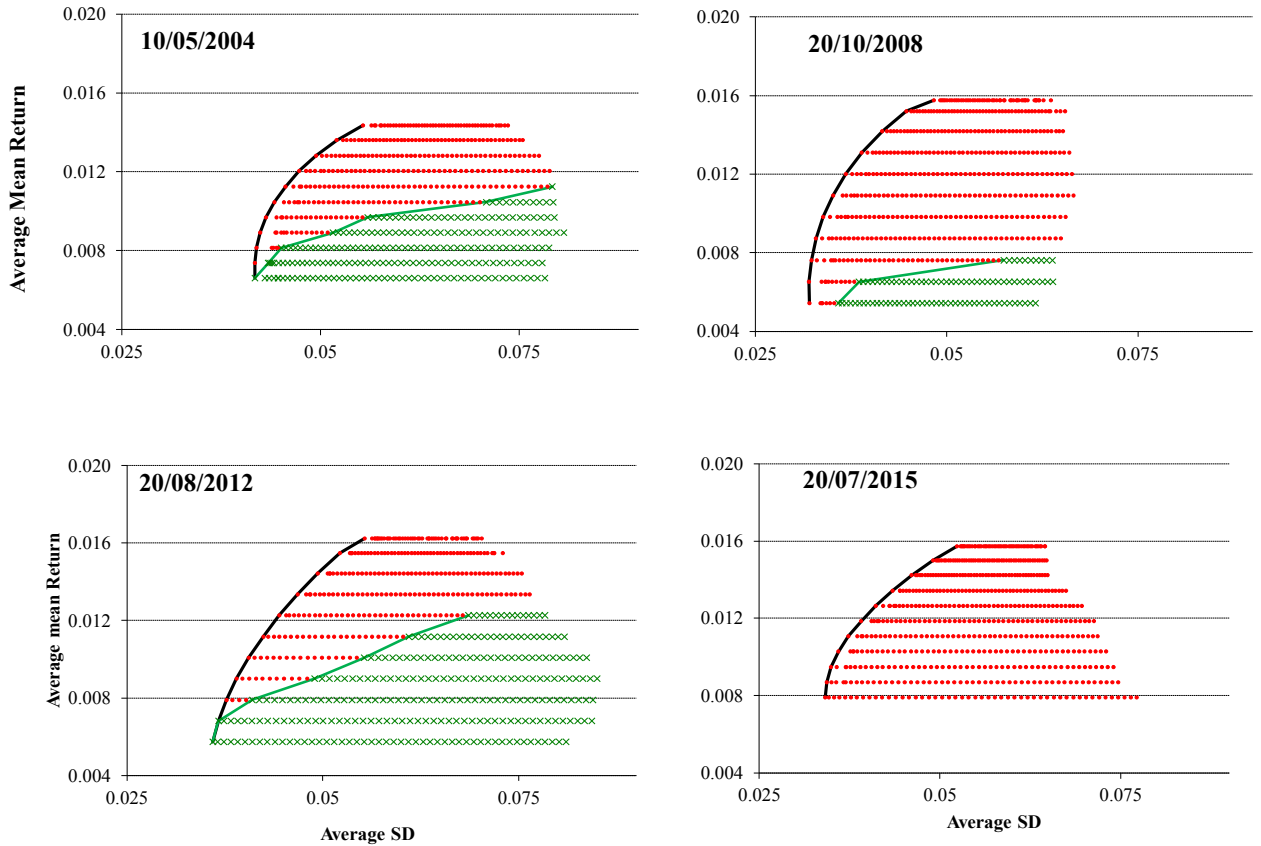

**Figure 4. Time series of the proportion of consistent portfolios from 30 Dow Jones stocks with the running 3 year maximum and (negative) minimum return on the DJ30 Index. The evaluation dates used in Figure 3 are also indicated.**

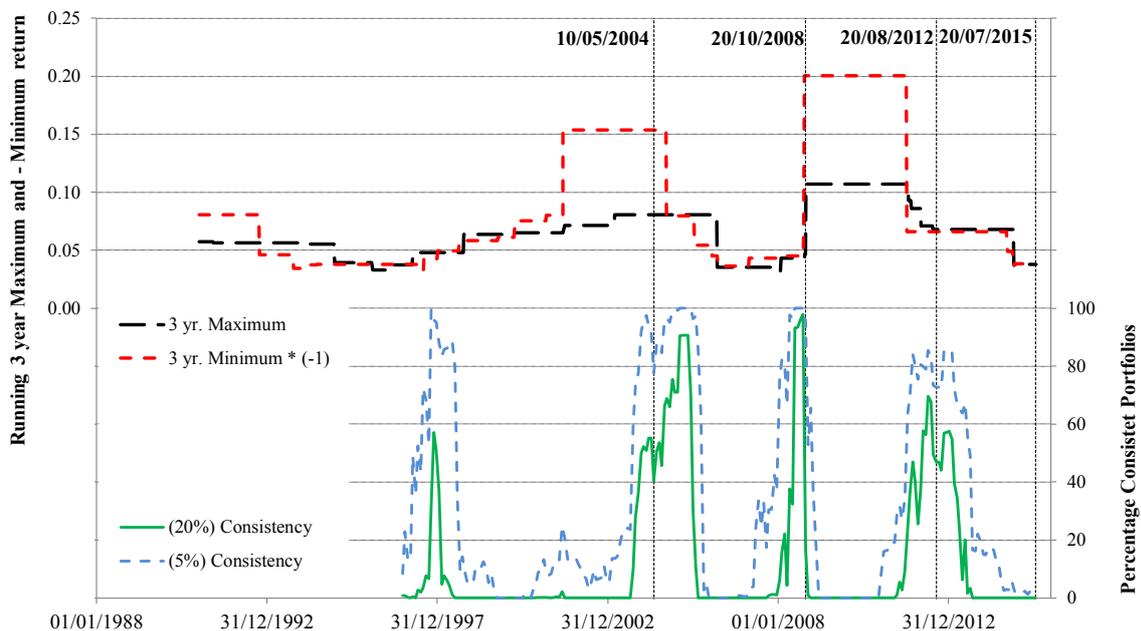

To summarise the fluctuation over time of the size of the consistency region, in Figure 4 we show a time series of the proportion of the 550 portfolios shown in the plots that are consistent using both a 20% and 5% critical value for the Berkowitz statistic. Considering acceptance at the 20% critical value, there are four periods from 1996 to 2015 where the consistency region includes a fifth or more of the efficient and dominated portfolios. During these periods, the plots on a particular date will resemble the first three plots in Figure 3. In between these periods there are runs of dates where there are few or no consistent portfolios, where the plots on a particular date resemble the fourth plot in Figure 3. Considering acceptance at the more lenient 5% critical value, the proportion of consistent portfolios is, by definition, higher but shows the same pattern. The cause of these fluctuations in the size of the consistency region is obviously time dependence in the data. As an aid to focussing more clearly on the more important aspects of time dependence, we show the maximum weekly return on the Dow Jones 30 Index over the previous three years and the negative of its minimum return. The interval of three years is equivalent to $K = 39$ in the computation of the Berkowitz statistic. There is a tendency for the consistency region to shrink sharply when the running maximum or (negative) minimum return increases. This reaction is plausible as the Berkowitz statistics detect the increase in volatility (indicated by the running maximum or minimum return) that will decrease the accuracy of return and risk estimates. A less attractive property is a tendency for the consistency region to increase in size after a drop in the running maximum or (negative) minimum return. The underlying reason here is that the unexpected returns that caused the decrease in the consistency region have dropped out of the calculation of the Berkowitz statistics. Whereas the response to an increase in volatility when it occurs is desirable, the response to the increase in volatility dropping out of the range of the Berkowitz statistic is undesirable. To demonstrate the improvement in the sensitivity of the extended Berkowitz statistic (using (4)) compared to standard version (using (2)), the sizes of the consistency regions (denoted *ewma*) are plotted in Figure 5, using a 20% critical value and a discount factor of $\gamma = 0.94$. If there is a change in volatility, the *ewma* consistency region decreases at a similar point and rate to the conventional consistency region. However, the *ewma* consistency region increases in advance of the conventional consistency region because the destabilising returns are discounted gradually compared to a stepwise change in the conventional statistic.

**Figure 5. A comparison of time series of the proportion of consistent portfolios from 30 Dow Jones stocks using the *ewma* (with discount factor of 0.94) Berkowitz statistic to the conventional Berkowitz statistic (both using a 20% critical value).**

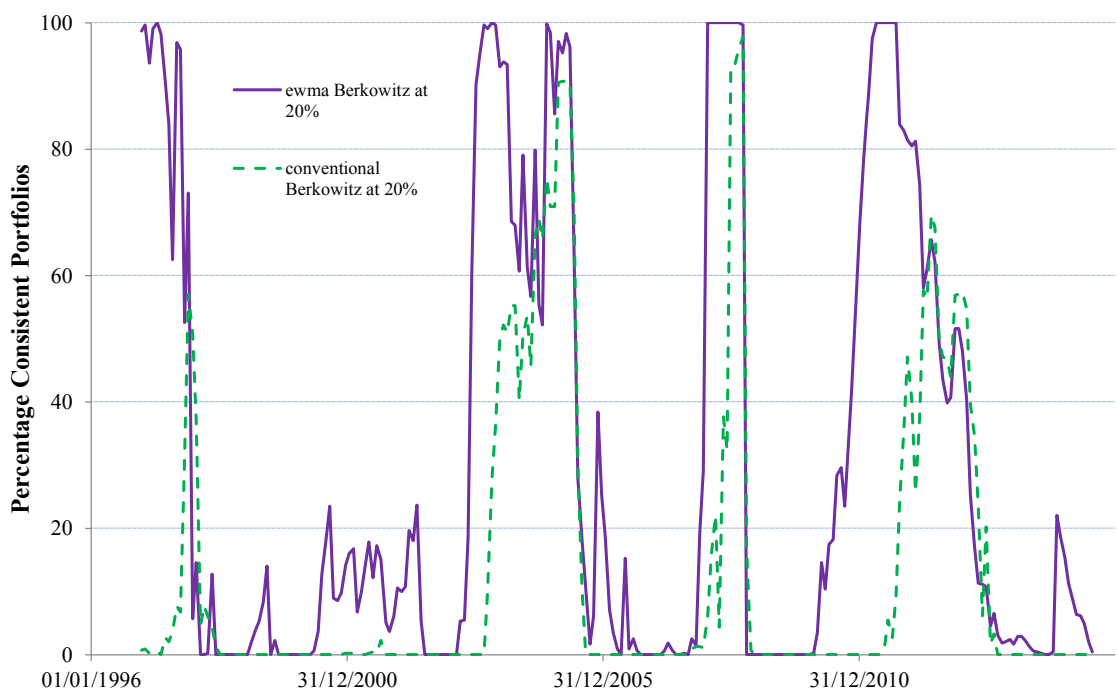

*7.2 Investment strategy implications of the consistency region*

To judge the worth of investing in portfolios that belong to the consistency region (or not) we use the in-sample Sharpe ratio of portfolio returns as a predictor of investment performance. The use of the Sharpe ratio for demonstrating differences in performance is a common practice[4], even when there is parameter or model uncertainty, see for example, Garlappi, Uppal and Wang (2007). Lo (2002) formulates the standard error of the ratio under a range of different conditions and Ledoit and Wolf (2008) suggest a robust hypothesis test using the ratio.

---

[4] Eling and Schuhmacher (2007) compare hedge fund rankings based on a range of performance measures including the Sharpe ratio and found no difference in the rankings. Although there are exceptions, it is widely reported in the performance measurement literature that many well established performance measures result in the same rankings as the Sharpe ratio.

Using the in-sample Sharpe ratio as the criterion for portfolio choice, we compare two strategies:

- A: for each period, invest in the portfolio with the highest Sharpe ratio;
- B: for each period, invest in the consistent portfolio with the highest Sharpe ratio. If there are no consistent portfolios, stay in cash and receive zero return.

The results of the strategy comparison are shown in Table 6 for a range of values of $\gamma$. We use the first 200 investment periods to measure in-sample performance. For $\gamma = 0.92, 0.94, 0.96, 1.00$, Strategy B produces both higher returns and better Sharpe ratios than Strategy A. Thus, for out-of-sample performance, we would choose $\gamma = 0.94$ as this value produces the highest returns. For the out-of-sample investment periods, this choice gives the second highest returns, but all values of $\gamma$ considered produce higher means than Strategy A.

**Table 6. Summary statistics for strategies A and B over a test period of 200 observations and an evaluation of 43 observations**

| Evaluation Period | | 1 - 200 | | | 201 - 243 | | |
|---|---|---|---|---|---|---|---|
| | | 23/12/1996 to 02/04/2012 | | | 30/04/2012 to 20/07/2015 | | |
| | | Mean | SD | Sharpe | Mean | SD | Sharpe |
| Strategy A | | 0.0043 | 0.0496 | 0.0868 | 0.0055 | 0.0306 | 0.1792 |
| Strategy B | 0.90 | 0.0038 | 0.0444 | 0.0857 | 0.0070 | 0.0269 | 0.2609 |
| | 0.92 | 0.0049 | 0.0424 | 0.1155 | 0.0061 | 0.0257 | 0.2395 |
| Discount | *0.94* | *0.0058* | *0.0416* | *0.1386* | *0.0070* | *0.0251* | *0.2791* |
| Factor | 0.96 | 0.0054 | 0.0416 | 0.1290 | 0.0078 | 0.0259 | 0.3017 |
| | 0.98 | 0.0037 | 0.0463 | 0.0790 | 0.0061 | 0.0270 | 0.2244 |
| | 1.00 | 0.0050 | 0.0441 | 0.1131 | 0.0059 | 0.0280 | 0.2109 |

## 8. Conclusions

To overcome the optimistically biased prediction of the ex-ante efficient frontier, we considered two approaches. Firstly, we introduced the concept of the consistency region in risk-expected return space, where a portfolio's in-sample performance is a reliable predictor of its out-of-sample return. The consistency region is bounded above by the consistency frontier. Inherent in this concept is the use and development of the Berkowitz statistic to improve its sensitivity to changes in the data generating process. Secondly, we extended ex-post efficient set mathematics to estimate the ex-post efficient frontier.

In an empirical validation using multivariate normal data with constant parameters, we find that the consistency region behaves intuitively reasonably, for short estimation periods the consistency region is composed of dominated portfolios with higher standard deviations than those on the efficient frontier with the same mean returns. As the estimation period is increased, the consistency region increases, gradually moving towards, and up, the efficient frontier. In this validation exercise, we find that the positions of the consistency frontier and the ex-post frontier are similar in that differences in portfolio expected returns and volatilities are generally small. In particular, the newly developed estimation of the ex-post frontier, Method 1, closely replicates the values of CVaR along the consistency frontier. Although the consistency of the ex-post frontier decreases for higher expected returns, when the ex-post frontier is consistent it falls just inside the consistency frontier. The simulation method may be extended to deal with more general models for the distribution of returns and forecasts. In the literature review we discussed two paths, Path One - enhancing parameter estimation, Path Two - mitigating the influence of estimation errors by adjusting constraints and objective functions. In this context, this validation exercise shows that the proposed methodology contributes to both paths and acts as a link between them.

For use with real data, we developed the Berkowitz statistic to make it more reactive the changing dynamics of asset returns. Used with actual data, the behaviour of the consistency region reflects these changing dynamics; for the same size estimation period, the size of the consistency region fluctuates. The proportion of portfolios that are consistent varies between near unity and zero. A comparison of investment strategies based on choosing maximum Sharpe ratio portfolios ex-ante on either the efficient frontier or the consistency frontier showed that the use of the consistency frontier led to a noticeably greater Sharpe ratio ex-post.

A central idea underlying this study is the use of density forecasting accuracy as a criterion for portfolio selection in the apparently sub-optimal space dominated by the efficient frontier. This approach could be applied elsewhere, for example in optimisation relative to a benchmark such as portfolio selection for index tracking. The methodology could be implemented with different approaches to the grid construction and other estimates of mean returns and the covariance matrix.

# Appendix

## A. Calibration of the critical values for the Berkowitz statistic

The result that the Berkowitz statistic is chi-squared is an asymptotic one. In our experiment, the sample size is not large and the use of overlapping data for the estimation of the *cdf* may lead to departures from a chi-squared distribution. To achieve appropriate critical values for the experiment, we use a simulation procedure where a time series of observations of identically and independently distributed normal random variables was generated. Using this series a set of *K* out-of-sample returns over *H* periods was calculated and their corresponding empirical *cdf*s calculated from the previous overlapping returns replicates the estimation methodology described above. For different values of *M* and *K*, the values of the Berkowitz statistic under the null hypothesis were simulated 20000 times and the 80, 85, 90 and 95 percentiles of the statistic were found. This exercise was performed five times and the average values used. For most combinations of *M* and *K*, the empirically derived critical value was less than the tabulated value. The ratios of the empirical critical value to the tabulated value for the 80 percentile and a chi-squared with two degrees of freedom are shown for a range of values for *M* and *K* in Table A1. The empirical critical values for the appropriate values of *M* and *K* are used throughout our analysis.

In order to test the sensitivity of the Berkowitz statistic to departures from the null hypothesis, using the empirically derived critical values, we repeat the simulation experiment with the standard deviation of the out-of-sample returns multiplied by $\theta$. We then observe the power of the Berkowitz statistic to reject the null hypothesis that the out-of-sample return is generated by the same process as the in-sample returns for different values of $\theta$. As an example, we show the power of the test for $M = 312$ (in-sample returns), $K = 39$ (forecast origins) and $\theta$ taking on a range of values from 0.5 to 1.5 in Figure A1. The power of the Berkowitz statistic to reject the null hypothesis at levels of significance of 20%, 10% and 5%, is plotted. For $\theta = 1$, the probabilities of rejection of a correct hypothesis are at their expected values for 20%, 10% and 5%. The probability of detection of a departure from the identified distribution increases with $|\theta - 1|$. For the 20% test, the probability of rejection is 60% for $\theta < 0.8$ and $\theta > 1.2$. In other words, if the standard deviation of the out-of-sample return differs by 20% or more from its in-sample value, there is a probability of 60% or more of detecting this (and hence rejecting the null hypothesis) using a test with a 20% significance level.

**Figure A1. The power of the Berkowitz statistic to reject the null hypothesis at levels of significance of 20%, 10% and 5% for changes in the standard deviation of out-of-sample returns (using M = 312 and K = 39)**

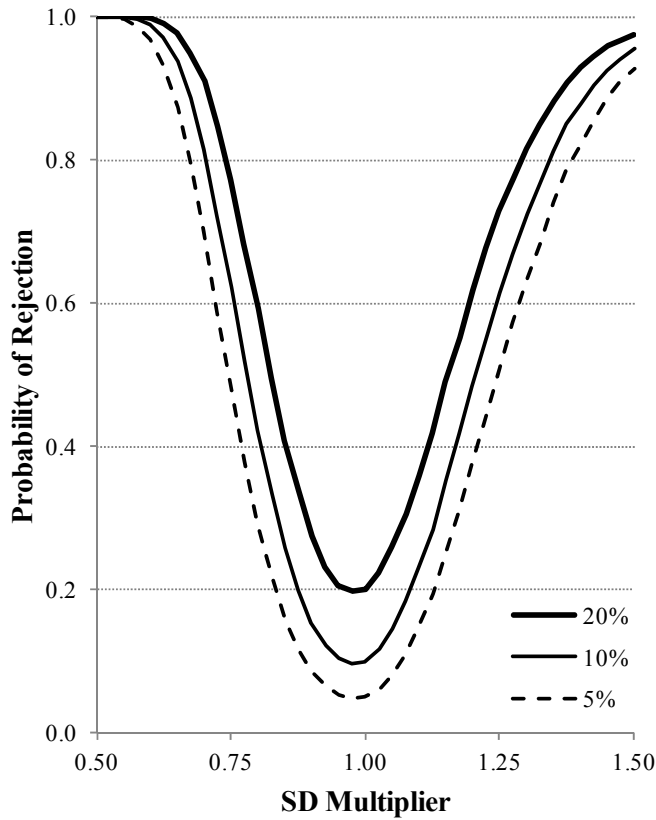

**Table A1. The ratio of the simulation derived critical values to the tabulated value $\chi^2_{2,20\%} = 3.2189$ for different values of M and K.**

|  |  | Values of K | | | | | | |
|---|---|---|---|---|---|---|---|---|
|  |  | 26 | 39 | 52 | 65 | 78 | 91 | 104 |
|  | 52 | 0.78 | 0.87 | 0.99 | 1.06 | 1.22 | 1.36 | 1.54 |
|  | 78 | 0.84 | 0.79 | 0.74 | 0.83 | 0.87 | 0.93 | 0.98 |
|  | 104 | 0.84 | 0.75 | 0.74 | 0.81 | 0.78 | 0.79 | 0.84 |
|  | 130 | 0.89 | 0.81 | 0.69 | 0.67 | 0.72 | 0.69 | 0.70 |
|  | 156 | 0.96 | 0.82 | 0.71 | 0.65 | 0.61 | 0.66 | 0.63 |
|  | 182 | 0.98 | 0.88 | 0.76 | 0.69 | 0.64 | 0.58 | 0.58 |
| **Values of M** | 208 | 1.00 | 0.88 | 0.77 | 0.66 | 0.63 | 0.61 | 0.56 |
|  | 234 | 0.93 | 0.85 | 0.83 | 0.69 | 0.63 | 0.62 | 0.56 |
|  | 260 | 0.94 | 0.93 | 0.86 | 0.73 | 0.71 | 0.62 | 0.57 |
|  | 286 | 0.98 | 0.93 | 0.85 | 0.82 | 0.69 | 0.62 | 0.58 |
|  | 312 | 0.96 | 0.89 | 0.86 | 0.82 | 0.73 | 0.66 | 0.61 |
|  | 338 | 0.91 | 0.90 | 0.85 | 0.82 | 0.79 | 0.70 | 0.64 |
|  | 364 | 1.02 | 0.96 | 0.93 | 0.83 | 0.83 | 0.72 | 0.67 |
|  | 390 | 0.93 | 0.93 | 0.88 | 0.88 | 0.80 | 0.77 | 0.71 |
|  | 416 | 0.91 | 0.90 | 0.89 | 0.82 | 0.81 | 0.75 | 0.73 |

**References**

Adcock, C. J., M. C. Cortez, M. R. Armada and F. Silva, 2012, Time Varying Betas and The Unconditional Distribution of Asset Returns, *Quantitative Finance,* 12, no. 6:951-967.

Adcock, C. J., 2013, Ex Post Efficient Set Mathematics, *The Journal of Mathematical Finance*, 3, no.1A: 201-210.

Adcock, C.J., 2014, Mean–variance–skewness efficient surfaces, Stein's lemma and the multivariate extended skew-Student distribution, *European Journal of Operational Research*, 234(2) 392-401.

Bao, Y., T-H. Lee and B. Saltoglu, 2007, Comparing density forecast models. *Journal of Forecasting*, 26(3), 203-225.

Barndorff-Nielsen, O.E., 1997, Normal inverse Gaussian distributions and stochastic volatility modelling, *Scandinavian Journal of Statistics*, 24(1), 1-13.

Berkowitz, J., 2001, Testing density forecasts, with applications to risk management. *Journal of Business & Economic Statistics*, 19(4), 465-474.

Black, F., and R. Litterman, 1991, Asset equilibrium: Combining investor views with market equilibrium, *Journal of Fixed Income*, 1, 7–18.

Blattberg, R.C. and N.J. Gonedes, 1974, A comparison of the stable and Student distributions as statistical models for stock prices, *Journal of Business*, 47(2), 244-280.

Bloomfield T., R. Leftwich and J.B. Long, 1977, Portfolio strategies and performance. *Journal of Financial Economics*, 5(2), 201-218.

Broadie, M., 1993, Computing efficient frontiers using estimated parameters. *Annals of Operations Research*, 45(1), 21-58.

Chan, L.K.C., J. Karceski and J. Lakonishok, 1999, On portfolio optimization: forecasting covariances and choosing the risk model. *The Review of Financial Studies*, 12(5), 937-974.

Chopra, V.K., C.R. Hensel and A.L. Turner, 1993, Massaging mean-variance inputs: returns from alternative global investment strategies in the 1980s. *Management Science*, 39(7), 845-855.

Corradi V. and N.R. Swanson, 2006, Predictive density and conditional confidence interval accuracy tests, *Journal of Econometrics,* 135(1-2), 187-228.

DeMiguel, V., L. Garlappi, F.J. Nogales and R. Uppal, 2009, A generalized approach to portfolio optimization: improving performance by constraining portfolio norms, *Management Science*, 55(5), 798-812.

DeMiguel, V., L. Garlappi, and R. Uppal, 2009, Optimal versus naive diversification: How inefficient is the 1/N portfolio strategy?, *The Review of Financial Studies,* 22(5), 1915-1953.

Diebold F.X., T.A. Gunther and A.S. Tay, 1998, Evaluating density forecasts with applications to financial risk management, *International Economic Review*, 39(4), 863-883.

Disatnik, D.J. and S. Benninga, 2007, Shrinking the covariance matrix, *Journal of Portfolio Management*, 33(4), 55-63.

M Eling, M. and Schuhmacher, F., 2007, Does the choice of performance measure influence the evaluation of hedge funds? *Journal of Banking & Finance*, 31(9), 2632-2647.

Fabozzi, F. J., P. N. Kolm, D. Pachamanova and S. M. Focardi , 2007. Robust portfolio optimization and management. New York: Wiley

Fabozzi, F.J., D.S. Huang, and G.F. Zhou, 2010, Robust portfolios: contributions from operations research and finance, *Annals of Operations Research*, 176(1), 191-220.

Frankfurter, G.M., H.E. Phillips and J.P. Seagle, 1971, Portfolio selection, the effects of uncertain means, variances, and covariances. *Journal of Financial and Quantitative Analysis*, 6(5), 1251-1262.

Garlappi, L., R. Uppal and T. Wang, 2007, Portfolio selection with parameter and model uncertainty: A multi-prior approach, *The Review of Financial Studies*, 20(1), 41-81.

Hong, Y.M. and H.T. Li, 2005, Nonparametric specification testing for continuous-time models with applications to term structure of interest rates, *The Review of Financial Studies*, 18(1), 37-84.

Hong Y.M., H.T. Li and F. Zhao, 2007, Can the random walk model be beaten in out-of-sample density forecasts? Evidence from intraday foreign exchange rates, *Journal of Econometrics*, 141(2), 736-776.

Hwang, I., S. Xu, and F. In, 2018, Naive versus optimal diversification: Tail risk and performance, *European Journal of Operational Research*, 265, 372–388.

Jagannathan, R. and T.S. Ma, 2003. Risk reduction in large portfolios: Why imposing the wrong constraints helps. *Journal of Finance*, 58(4), 1651-1683.

James, W. and C. Stein, 1961, Estimation with quadratic loss. *Proceedings of the 4$^{th}$ Berkeley Symposium on Probability and Statistics 1*. Berkeley: University of California Press, 361-379.

Jobson, J.D. and B. Korkie, 1980, Estimation for Markowitz efficient portfolios, *Journal of the American Statistical Association*, 75(371), 544-554.

Jorion, P., 1985, International portfolio diversification with estimation risk. *Journal of Business*, 58(3), 259-278.

Jorion, P., 1986, Bayes-Stein estimation for portfolio analysis. *The Journal of Financial and Quantitative Analysis*, 21(3), 279-292.

Jorion, P., 1991, Bayesian and CAPM estimators of the means: Implications for portfolio selection, *Journal of Banking and Finance*, 15(3), 717-727.

Kapsos, M., N. Christofides, and B. Rustem, 2014, Worst-case robust Omega ratio, *European Journal of Operational Research*, 234, 499–507.

Kan, R. and D. R. Smith, 2008, The Distribution of the Sample Minimum-Variance Frontier, *Management Science*, 54, no. 7:1364-1360

Kim, J., W. Kim, and F. Fabozzi, 2014, Recent developments in robust portfolios with a worst-case approach. *Journal of Optimization Theory and Applications*, 161(1):103–121.

Kolm, P.N., Tütüncü, R. and Fabozzi, F.J., 2014, 60 Years of portfolio optimization: Practical challenges and current trends, *European Journal of Operational Research*, 234 356–371

Kon, S.J., 1984, Models of stock returns - a comparison, *Journal of Finance*, 39( 1), 147-165.

Ledoit, O. and M. Wolf, 2003, Improved estimation of the covariance matrix of stock returns with an application to portfolio selection. *Journal of Empirical Finance,* 10(5), 603-621.

Ledoit, O. and M. Wolf, 2008, Robust performance hypothesis testing with the Sharpe ratio, *Journal of Empirical Finance*, 15(5), 850-859.

Lo, A.W., 2002, The statistics of Sharpe ratios, *Financial Analysts Journal*, 58(4), 36-52.

Maillard, S., T. Roncalli and J. Teïletche, 2010, The Properties of Equally Weighted Risk Contribution Portfolios, *The Journal of Portfolio Management*, 36 (4) 60-70.

Markowitz, H., 1952. Portfolio selection. *Journal of Finance*, 7(1), 77–91.

Markowitz, H., 2014, Mean–variance approximations to expected utility. *European Journal of Operational Research*, 234, 346–355.

Mathai, A.M. and S. B. Prevost, 1992, Quadratic Forms in Random Variables, Heidelberg, Springer.

Meade, N., 2010, Oil prices - Brownian motion or mean reversion? A study using a one year ahead density forecast criterion, *Energy Economics*, 32(6), 1485-1498.

Michaud, R.O., 1989, The Markowitz optimization enigma: is 'optimized' optimal? *Financial Analysts Journal*, 45(1), 31-42.

Michaud, R. and R. Michaud, 2007, Estimation Error and Portfolio Optimization: A Resampling Solution, *Journal of Investment Management*, 6 (1), 8 – 28.

Mitchell, J. and S.G. Hall, 2005, Evaluating, comparing and combining density forecasts using the KLIC with an application to the Bank of England and NIESR 'fan' charts of inflation. *Oxford Bulletin of Economics and Statistics*, 67, Supplement s1, 995-1033.

Palczewski, A. and J. Palczewski, 2014, Theoretical and empirical estimates of mean–variance portfolio sensitivity, *European Journal of Operational Research*, 234 (2), 402-410.

Polson, N.G. and B.V. Tew, 2000, Bayesian portfolio selection: An empirical analysis of the S&P 500 index 1970-1996, *Journal of Business & Economic Statistics*, 18(2), 164-173.

Rockafellar, R.T. and S. Uryasev, 2000, Optimization of conditional value-at-risk, *Journal of Risk*, 2, 21-41.

Simões, G., M. McDonald, S. Williams, D. Fenn and R. Hauser, 2018, Relative Robust Portfolio Optimization with benchmark regret, *Quantitative Finance*, 18 (12), 1991-2003.

Spellucci, P., 1998a, An SQP method for general nonlinear programs using only equality constrained subproblems, *Mathematical Programming*, 82, 413 - 448.

Spellucci, P., 1998b, A new technique for inconsistent problems in the SQP method, *Mathematical Methods of Operational Research*, 47, 355 - 400.

Tu, J. and G.F. Zhou, 2004, Data-generating process uncertainty: What difference does it make in portfolio decisions? *Journal of Financial Economics*, 72(2), 385-421.

Tütüncü, R.H. and M. Koenig, 2004, Robust asset allocation. *Annals of Operations Research*, 132(1), 157 - 187.

Wang, Z.Y., 2005, A shrinkage approach to model uncertainty and asset allocation, *The Review of Financial Studies*, 18(2), 673-705.

Yu, J-R., W-J. P. Chiou, W-Y. Lee, and T-Y. Chuang, 2019, Realized performance of robust portfolios: worst-case Omega vs CVaR related models, *Computers and Operations Research*, 104, 239-255.